## Modeling the Health Expenditure in Japan, 2011. A Healthy Life Years Lost Methodology

**Christos H Skiadas[1] and Charilaos Skiadas[2]**

[1] ManLab, Technical University of Crete, Chania, Crete, Greece
(E-mail: skiadas@cmsim.net )
[2] Department of Mathematics and Computer Science, Hanover College, Indiana, USA
(E-mail: skiadas@hanover.edu )

## Abstract

The Healthy Life Years Lost Methodology (HLYL) is introduced to model and estimate the Health Expenditure in Japan in 2011. The HLYL theory and estimation methods are presented in our books in the Springer Series on Demographic Methods and Population Analysis vol. 45 and 46 titled: "Exploring the Health State of a Population by Dynamic Modeling Methods" and "Demography and Health Issues: Population Aging, Mortality and Data Analysis". Special applications appear in Chapters of these books as in "The Health-Mortality Approach in Estimating the Healthy Life Years Lost Compared to the Global Burden of Disease Studies and Applications in World, USA and Japan" and in "Estimation of the Healthy Life Expectancy in Italy Through a Simple Model Based on Mortality Rate" by Skiadas and Arezzo. Here further to present the main part of the methodology with more details and illustrations, we develop and extend a life table important to estimate the healthy life years lost along with the fitting to the health expenditure in the related case. The application results are quite promising and important to support decision makers and health agencies with a powerful tool to improve the health expenditure allocation and the future predictions.

## 1. Introduction

Life Tables have dominated the quantitative and qualitative demography issues for the last 4 centuries. Following the works of John Graunt (1676) and Edmond Halley (1693), life tables became the standard measure of several demographic parameters. Later on, models as the Gompertz (1825) and Makeham (1860) came to support and straighten the use of mortality models in demography and insurance while giving rise to various approaches and studies in probability and statistics.

However, a further exploration of the hidden properties and functions in the classical life tables is needed. It was already demonstrated in several publications in the last decades and more recently with few publications (see Janssen and Skiadas (1995), Skiadas, C.H. and Skiadas, C. (2010, 2014, 2015, 2018 a,b,c,d), Skiadas and Arezzo (2018), Strehler & Mildvan (1960), Sullivan (1966, 1971), Torrance (1976), Chia. and Peng Loh (2018)). The main task was the use of appropriate transformations in the life tables in order to extract valuable information. To this end, the estimation of the life years lost to disabilities and health deterioration is of particular importance. Our works on estimating this important period of the life time are in parallel to the other approaches developed under the Global Burden of Disease (GBD) terminology (WHO, 2014). The latter is a very complicated and heavy international methodology important to estimate the impact of every particular kind of disease and disability on the health state of the population and consequently on the health expenditure in a country. The results are produced as Healthy Life Years Lost (HLYL) or as the estimated Healthy Life Expectancy (HLE) that is the Life Expectancy (LE) minus the HLYL. The GBD results regarding the particular kind of diseases and disabilities are the most important part of these studies. A complicated methodology weights and interconnects the particular kinds of disabilities and diseases in a final number for the HLE provided by agencies as the World Health Organization (WHO) as HALE and health statistical bureaus under several terminologies. Following our estimates, the HLYL can be found directly from the life tables with a relative simple methodology presented in recent publications (Skiadas & Skiadas (2018 a,b,c,d). The method is based on developing a Survival vs Mortality diagram and do direct estimates based on mx or qx as termed in the well-developed life tables of the Human Mortality Database (HMD). To further straighten the validity and importance of this methodology we also have provided a parallel HLYL estimates based on the Gompertz (1825) and Weibull (1951) models. A brief presentation of the related methodology is presented here.

The main task of this paper is to explore on how the healthy life years lost estimated can be used to calculate the Health Expenditure in a Country. Clearly the up to now estimates of life expectancy, healthy life expectancy and healthy life years lost are not easily applied to the health expenditure calculations. Instead our approach of the HLYL estimates is directly used in the health expenditure calculations as we present in the following.

## 2. The HLYL Estimation Method

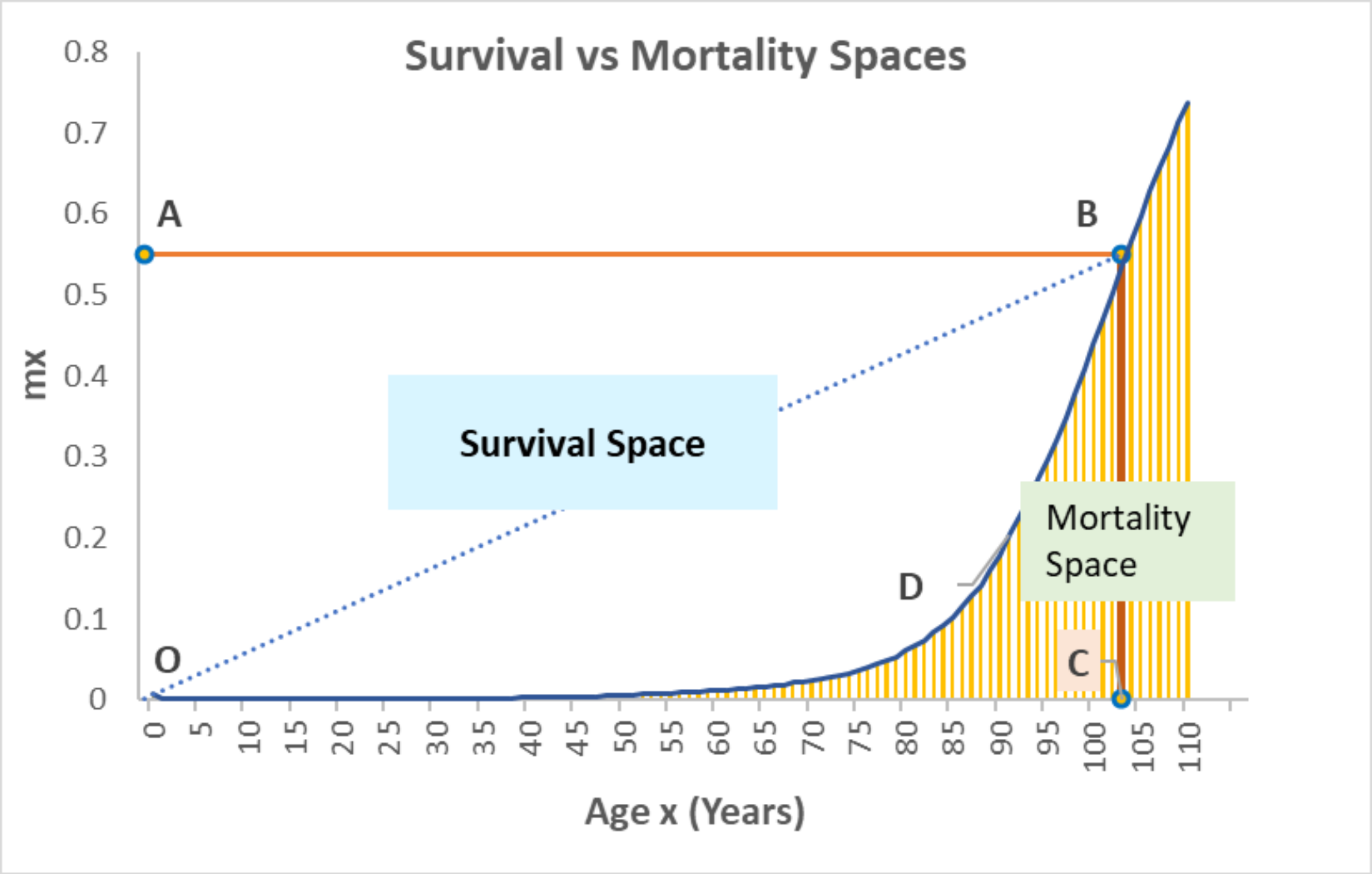

Fig. 1. The Survival – Mortality spaces.

The simplest form to express mortality mx in a population at age x is by estimating the fraction Death(D)/Population(P) that is mx=D/P. Accordingly the above graph is formulated with mx as the blue exponential curve whereas the horizontal axis express years of age. The main forms of Life Tables start with mx and in the following estimate qx and the survival forms of the population. This methodology leads to the estimation of a probability measure termed as life expectancy at age x or life expectancy at birth when considering the total life time. There are several differences between the graph with the survival space above and the survival curves methodology. First of all, the vertical axis in the Survival-Mortality

Space (SMS) diagram is the probability mx. Instead in the survival diagram the vertical axis represent population (usually it starts from 100.000 in most life tables and gradually slow down until the end). By the SMS diagram we have probability spaces for both survival and mortality. For the age x the total space is (ABCOA) in the SMS diagrams that is (OC)x(OA)=x.mx. The mortality space (ODBCO) is the sum S(mx) and the survival space is (x.mx-S(mx)). Accordingly, the important measure of the Health State is simply the fraction (ABDOA)/(BCODB). This is to provide the form FHM*=(x.mx-S(mx))/S(mx)=x.mx/S(mx)-1. Simpler is to prefer the fraction FHM=(ABCOA)/(BCODB)=x.mx/S(mx) that will be also estimated from mx for every age x of the population.

The FHM is calculated by:

$$FHM(mx) = \lambda \frac{xmx}{\sum_1^x mx}$$

Note that a similar approach based on qx estimates in a life table can be estimated with the following formula:

$$FHM(qx) = \lambda \frac{xqx}{\sum_1^x qx}$$

In both cases the parameter λ has to be estimated. In the majority of applications λ=1 provides quite good results.

Both estimates are presented in the following figure 2 obtained from the Japan 2011 Extended Full Life Table (see also figure 7). Both methods show similar results until 70 years of age and then the mx estimates give higher values than the qx. In both cases a maximum is reached in the area from 95 -100 years of age followed by a decline in the remaining years. The estimated HLYL are 10.1 years for the mx based estimate and 9.5 years for the qx based estimate. As the life expectancy at birth for Japan 2011 is 82.7 years of age, the estimates for the health life expectancy are 72.6 and 73.2 years of age for the estimates based on mx and qx respectively. In the related figure 2 the healthy years lost estimated with the Healthy Life Expectancy method are also presented with a cyan line. The latter is mainly a smoothing like presentation of the HLYL estimates from mx and qx.

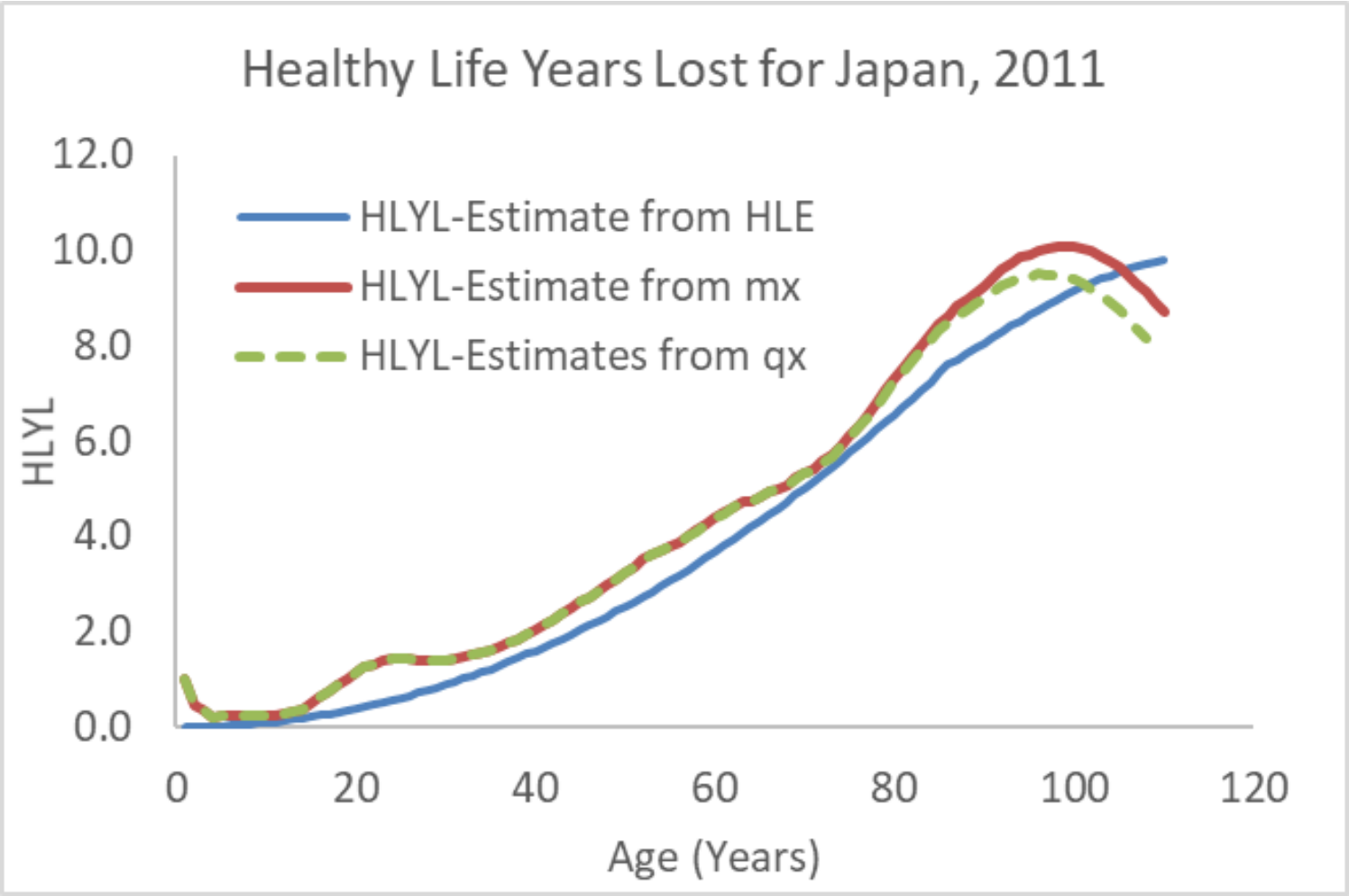

Fig. 2. Healthy life years lost estimated.

The Health/Mortality fraction has an increasing form for the main part of the lifetime until a maximum and then a decline. The fraction for Japan 2011 is similar until 70 years of age for both estimates from mx or qx. The very important point here is that the maximum points correspond to the maximum years of age lost to disability. We can easily observe this important future by considering a linear form for mortality mx=ax. This is the simplest case of drawing a linear line from O to B in the graph above. The resulting fraction is 2 whereas is 1 if we select the fraction (ABDOA)/(BCODB). Following the previous discussion, the healthy life years lost to disability (HLYL) are 1 with the last notation and 2 when considering the total space vs the mortality space. The latter higher by 1 from the simple fraction provides results similar to those estimated by the World Health Organization. After that, the only we have is to remove this estimate from the life expectancy at birth to find the Healthy Life Expectancy.

Using the 2011 Life Table data-set for both (male and female) from the Japan official website we found the following Table of the estimates. Included are the estimates from the Gompertz and Weibull models following our methodology presented in our books. From the same reference are also the direct estimates based on mx and qx provided from the life table. The healthy life years lost per year of age and the healthy life expectancy are also included. The full life table with estimates is presented in the end of this paper in Figure 7.

| **Japan-b-2011** | | | | | | | | | | | **Healthy Life Years Lost (HLYL)** | | | | |
|---|---|---|---|---|---|---|---|---|---|---|---|---|---|---|---|
| **Estimates of HLYL and HLE (C H Skiadas and C Skiadas, 2016).** Add data in this Excel file and do not change the supporting files Gompertz and Weibull as they run automatically | | | | | | | | | | | **HLYL Gompertz** | **HLYL Weibull** | **HLYL from mx** | **HLYL from qx** | **HLYL Average** |
| **Manually complete the blue Excel part only for mx or add all the Life Table from the Human Mortality Database, HMD (recommended)** | | | | | | | | | | **Years of age** | **9.87** | **9.89** | **10.09** | **9.50** | **9.84** |
| Year | Age | mx | qx | ax | lx | dx | Lx | Tx | ex | **HLYL** | **HLE** | | | | |
| 2011 | 0 | 0.00237 | 0.00237 | 0.21 | 100000 | 237 | 99813 | 8271181 | 82.71 | **9.8** | **72.87** | 0.00237 | 0.00237 | 0.00237 | 0.00237 |
| 2011 | 1 | 0.00042 | 0.00042 | 0.5 | 99763 | 42 | 99742 | 8171368 | 81.91 | **9.8** | **72.07** | 0.00279 | 0.00279 | 0.00042 | 0.00042 |
| 2011 | 2 | 0.00028 | 0.00028 | 0.5 | 99721 | 27 | 99707 | 8071626 | 80.94 | **9.8** | **71.10** | 0.00307 | 0.00307 | 0.00056 | 0.00056 |
| 2011 | 3 | 0.00023 | 0.00023 | 0.5 | 99694 | 23 | 99682 | 7971919 | 79.96 | **9.8** | **70.13** | 0.0033 | 0.0033 | 0.00069 | 0.00069 |
| 2011 | 4 | 0.00018 | 0.00018 | 0.5 | 99671 | 18 | 99662 | 7872236 | 78.98 | **9.8** | **69.16** | 0.00348 | 0.00348 | 0.00072 | 0.00072 |
| 2011 | 5 | 0.00016 | 0.00016 | 0.5 | 99653 | 16 | 99645 | 7772575 | 78 | **9.8** | **68.19** | 0.00364 | 0.00364 | 0.0008 | 0.0008 |
| 2011 | 6 | 0.00016 | 0.00016 | 0.5 | 99637 | 16 | 99629 | 7672930 | 77.01 | **9.8** | **67.21** | 0.0038 | 0.0038 | 0.00096 | 0.00096 |
| 2011 | 7 | 0.00014 | 0.00014 | 0.5 | 99621 | 14 | 99614 | 7573301 | 76.02 | **9.8** | **66.23** | 0.00394 | 0.00394 | 0.00098 | 0.00098 |
| 2011 | 8 | 0.00011 | 0.00011 | 0.5 | 99607 | 11 | 99602 | 7473687 | 75.03 | **9.8** | **65.26** | 0.00405 | 0.00405 | 0.00088 | 0.00088 |
| 2011 | 9 | 0.00012 | 0.00012 | 0.5 | 99596 | 12 | 99590 | 7374085 | 74.04 | **9.8** | **64.28** | 0.00417 | 0.00417 | 0.00108 | 0.00108 |
| 2011 | 10 | 0.00011 | 0.00011 | 0.5 | 99584 | 11 | 99579 | 7274494 | 73.05 | **9.7** | **63.31** | 0.00428 | 0.00428 | 0.0011 | 0.0011 |
| 2011 | 11 | 0.00011 | 0.00011 | 0.5 | 99573 | 11 | 99567 | 7174916 | 72.06 | **9.7** | **62.35** | 0.00439 | 0.00439 | 0.00121 | 0.00121 |
| 2011 | 12 | 0.0001 | 0.0001 | 0.5 | 99562 | 10 | 99556 | 7075348 | 71.07 | **9.7** | **61.38** | 0.00449 | 0.00449 | 0.0012 | 0.0012 |
| 2011 | 13 | 0.00013 | 0.00013 | 0.5 | 99551 | 13 | 99545 | 6975792 | 70.07 | **9.7** | **60.41** | 0.00462 | 0.00462 | 0.00169 | 0.00169 |
| 2011 | 14 | 0.00015 | 0.00015 | 0.5 | 99538 | 15 | 99530 | 6876247 | 69.08 | **9.6** | **59.45** | 0.00477 | 0.00477 | 0.0021 | 0.0021 |
| 2011 | 15 | 0.00018 | 0.00018 | 0.5 | 99523 | 18 | 99514 | 6776717 | 68.09 | **9.6** | **58.49** | 0.00495 | 0.00495 | 0.0027 | 0.0027 |
| 2011 | 16 | 0.00025 | 0.00025 | 0.5 | 99505 | 25 | 99493 | 6677203 | 67.1 | **9.6** | **57.53** | 0.0052 | 0.0052 | 0.004 | 0.004 |
| 2011 | 17 | 0.00029 | 0.00029 | 0.5 | 99480 | 29 | 99466 | 6577710 | 66.12 | **9.5** | **56.58** | 0.00549 | 0.00549 | 0.00493 | 0.00493 |
| 2011 | 18 | 0.00037 | 0.00037 | 0.5 | 99451 | 37 | 99432 | 6478245 | 65.14 | **9.5** | **55.64** | 0.00586 | 0.00586 | 0.00666 | 0.00666 |
| 2011 | 19 | 0.00036 | 0.00036 | 0.5 | 99414 | 36 | 99396 | 6378812 | 64.16 | **9.5** | **54.70** | 0.00622 | 0.00622 | 0.00684 | 0.00684 |
| 2011 | 20 | 0.0004 | 0.0004 | 0.5 | 99378 | 40 | 99358 | 6279416 | 63.19 | **9.4** | **53.77** | 0.00662 | 0.00662 | 0.008 | 0.008 |
| 2011 | 21 | 0.00045 | 0.00045 | 0.5 | 99339 | 45 | 99316 | 6180058 | 62.21 | **9.4** | **52.83** | 0.00707 | 0.00707 | 0.00945 | 0.00945 |
| 2011 | 22 | 0.00049 | 0.00049 | 0.5 | 99294 | 48 | 99270 | 6080741 | 61.24 | **9.3** | **51.91** | 0.00756 | 0.00756 | 0.01078 | 0.01078 |
| 2011 | 23 | 0.00052 | 0.00052 | 0.5 | 99246 | 51 | 99220 | 5981471 | 60.27 | **9.3** | **50.99** | 0.00808 | 0.00808 | 0.01196 | 0.01196 |
| 2011 | 24 | 0.00053 | 0.00053 | 0.5 | 99195 | 53 | 99168 | 5882251 | 59.3 | **9.2** | **50.07** | 0.00861 | 0.00861 | 0.01272 | 0.01272 |
| 2011 | 25 | 0.00051 | 0.00051 | 0.5 | 99141 | 50 | 99116 | 5783083 | 58.33 | **9.2** | **49.15** | 0.00912 | 0.00912 | 0.01275 | 0.01275 |
| 2011 | 26 | 0.00051 | 0.00051 | 0.5 | 99091 | 51 | 99066 | 5683967 | 57.36 | **9.1** | **48.23** | 0.00963 | 0.00963 | 0.01326 | 0.01326 |
| 2011 | 27 | 0.0005 | 0.0005 | 0.5 | 99040 | 50 | 99016 | 5584901 | 56.39 | **9.1** | **47.32** | 0.01013 | 0.01013 | 0.0135 | 0.0135 |
| 2011 | 28 | 0.00052 | 0.00052 | 0.5 | 98991 | 52 | 98965 | 5485886 | 55.42 | **9.0** | **46.40** | 0.01065 | 0.01065 | 0.01456 | 0.01456 |
| 2011 | 29 | 0.00056 | 0.00056 | 0.5 | 98939 | 56 | 98911 | 5386921 | 54.45 | **9.0** | **45.49** | 0.01121 | 0.01121 | 0.01624 | 0.01624 |
| 2011 | 30 | 0.00057 | 0.00057 | 0.5 | 98883 | 56 | 98855 | 5288010 | 53.48 | **8.9** | **44.59** | 0.01178 | 0.01178 | 0.0171 | 0.0171 |
| 2011 | 31 | 0.00058 | 0.00058 | 0.5 | 98827 | 57 | 98798 | 5189155 | 52.51 | **8.8** | **43.68** | 0.01236 | 0.01236 | 0.01798 | 0.01798 |
| 2011 | 32 | 0.00061 | 0.00061 | 0.5 | 98770 | 61 | 98740 | 5090356 | 51.54 | **8.8** | **42.78** | 0.01297 | 0.01297 | 0.01952 | 0.01952 |
| 2011 | 33 | 0.00061 | 0.00061 | 0.5 | 98709 | 60 | 98679 | 4991617 | 50.57 | **8.7** | **41.88** | 0.01358 | 0.01358 | 0.02013 | 0.02013 |

Fig. 3. The Heathy Life Years Lost and Healthy Life Expectancy estimates Extended Life Table form

The results obtained are from the Full Life Table for Japan. The related for the Abridged Life Table for 2011 are included in the following Table.

## 3. Modeling of the Health Expenditure in Japan, 2011

The results from two methods are similar. The abridged life tables are more useful as the provided health expenditure data for several countries follow an abridged life table population group schedule with estimates of five-year age groups. This is presented in the following figure 4 Extended Life Table for Japan 2011 along with the health expenditure estimates and comparisons with the real life data sets.

| Japan Life Table Data (period 5x, b) | | | | | | Ch Skiadas 18 February 2019 | | | |
|---|---|---|---|---|---|---|---|---|---|
| Year | Age | mx | Age x | mx | k.x.mx / sum(m1:mx) | Move Up | Japan Estimates | Japan Data | SSE |
| 2011 | | | | **0.01989** | **28.657** | **68** | | | |
| 2011 | 0--4 | 0.00265 | 2 | 0.01989 | 143.28 | 68 | 211 | 223 | 137 |
| 2011 | 5--9 | 0.00014 | 5 | 0.00014 | 1.40 | 68 | 69 | 123 | 2873 |
| 2011 | 10--14 | 0.00012 | 10 | 0.00012 | 2.05 | 68 | 70 | 86 | 254 |
| 2011 | 15-19 | 0.00029 | 15 | 0.00029 | 6.91 | 68 | 75 | 68 | 48 |
| 2011 | 20-24 | 0.00048 | 20 | 0.00048 | 14.46 | 68 | 82 | 71 | 131 |
| 2011 | 25-29 | 0.00052 | 25 | 0.00052 | 18.76 | 68 | 87 | 90 | 10 |
| 2011 | 30-34 | 0.00061 | 30 | 0.00061 | 25.36 | 68 | 93 | 106 | 160 |
| 2011 | 35-39 | 0.00083 | 35 | 0.00083 | 38.46 | 68 | 106 | 116 | 91 |
| 2011 | 40-44 | 0.00126 | 40 | 0.00126 | 62.81 | 68 | 131 | 132 | 1 |
| 2011 | 45-49 | 0.00189 | 45 | 0.00189 | 97.78 | 68 | 166 | 167 | 1 |
| 2011 | 50-54 | 0.00297 | 50 | 0.00297 | 152.59 | 68 | 221 | 210 | 112 |
| 2011 | 55-59 | 0.00445 | 55 | 0.00445 | 217.27 | 68 | 285 | 266 | 371 |
| 2011 | 60-64 | 0.00694 | 60 | 0.00694 | 305.25 | 68 | 373 | 352 | 452 |
| 2011 | 65-69 | 0.01021 | 65 | 0.01021 | 387.38 | 68 | 455 | 454 | 2 |
| 2011 | 70-74 | 0.01596 | 70 | 0.01596 | 494.71 | 68 | 563 | 614 | 2631 |
| 2011 | 75-79 | 0.02771 | 75 | 0.02771 | 648.58 | 68 | 717 | 771 | 2962 |
| 2011 | 80-84 | 0.05012 | 80 | 0.05012 | 815.65 | 68 | 884 | 910 | 694 |
| 2011 | 85-89 | 0.08928 | 85 | 0.08928 | 952.56 | 68 | 1021 | 1009 | 134 |
| 2011 | 90-94 | 0.15692 | 90 | 0.15692 | 1059.18 | 68 | 1127 | 1086 | 1696 |
| 2011 | 95-99 | 0.26304 | 95 | 0.26304 | 1118.64 | 68 | 1187 | 1167 | 386 |
| 2011 | 100+ | 0.40944 | 100 | 0.40944 | 1125.79 | 68 | 1194 | 1198 | 18 |
| | | | | | | | | **R2=** | **0.943** |
| | | | | | | | | **se=** | **25.04** |
| | | | | | | | 9117 | 9219 | 13164 |

Fig. 4. The extended abridged life table for health expenditure estimates

Health Expenditure Data from: Jennifer Friedman (2015) Health Expenditure per Capita: Application in Japan 2011, Public Lecture.

We recall that the FHM in Figure 4 is calculated by:

$$FHM(mx) = \frac{xmx}{\sum_{1}^{x} mx}$$

The estimates are based on the FHM multiplied by a parameter k (orange color) that is FHM=kx(mx)/sum(m1:mx). This parameter k=28.723 is estimated by a regression analysis so that to minimize the sum of squared errors between the data and the estimates. The first estimate for the mx* (green color) is selected from the data sets following the formula:

mx*= (D1-MIN(Data))/(sum(Data)-21*MIN(Data))=0.01989

where D1=223, Min(Data)=68 is the minimum data point and n=21 is the number of data points. The other points of the fifth column are equal to those provided for mx (column three). The sixth column includes the estimates based on FHM theory. Note that the formula for FHM takes the form:

$$FHM(mx) = k\frac{(x+2)mx}{\sum_1^x mx}$$

Where *x* is provided from column 4 in Figure 4.

The application to Japan 2011 showed that the FHM was a good approach. We have estimated the model with $R^2$=0.943. The standard error is se=25.04 corresponding to 5.7% of the mean of the data set.

Clearly our method indicates that there is no need to try to find any special model and estimate the parameters as all the needed data and modeling are included into the life tables. The only we have is to search further to find the hidden forms, to extract them and apply.

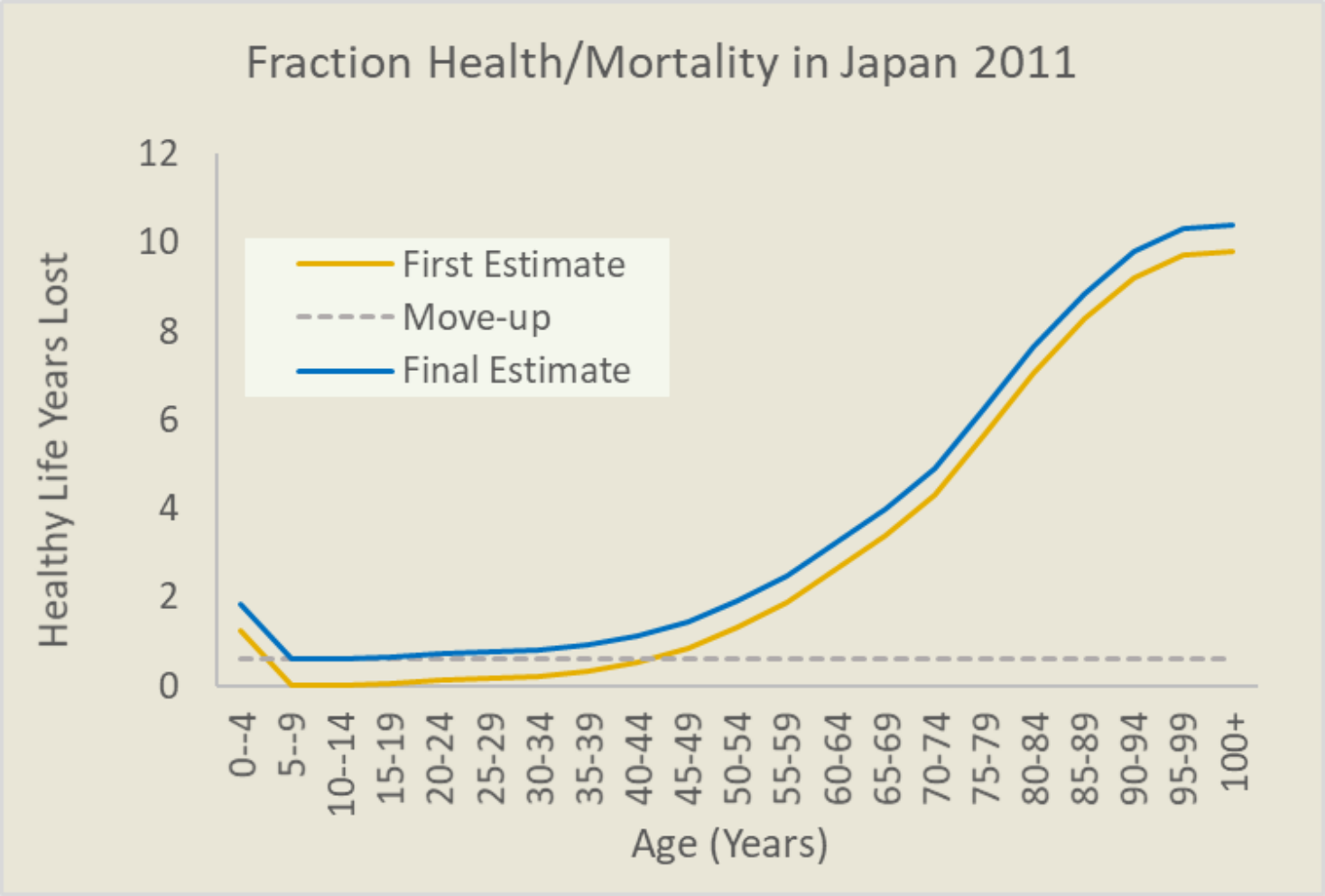

Fig. 5. FHM in Japan 2011 and move-up correction for the health expenditure.

In figure 5 we present the first estimate of the FHM done in the figure 4 life table followed by a move-up to the minimum level of the health expenditure data provided for Japan 2011. The Life Expectancy is 82.71 years according to the abridged life table for Japan in 2011 and the HLYL is 9.81 years of age. The estimated HLE is 72.90 years (Note that the HLE is higher when estimating with the qx method). Following the last graph, the health expenditure should be very low in the age interval from 5 to 15 years of age. However, the Health Systems have a minimum spending to be taken into account corresponding to an up movement of the FHM curve that is estimated in the Japan 2011 case to account at 5.68% of the maximum health expenditure per capita.

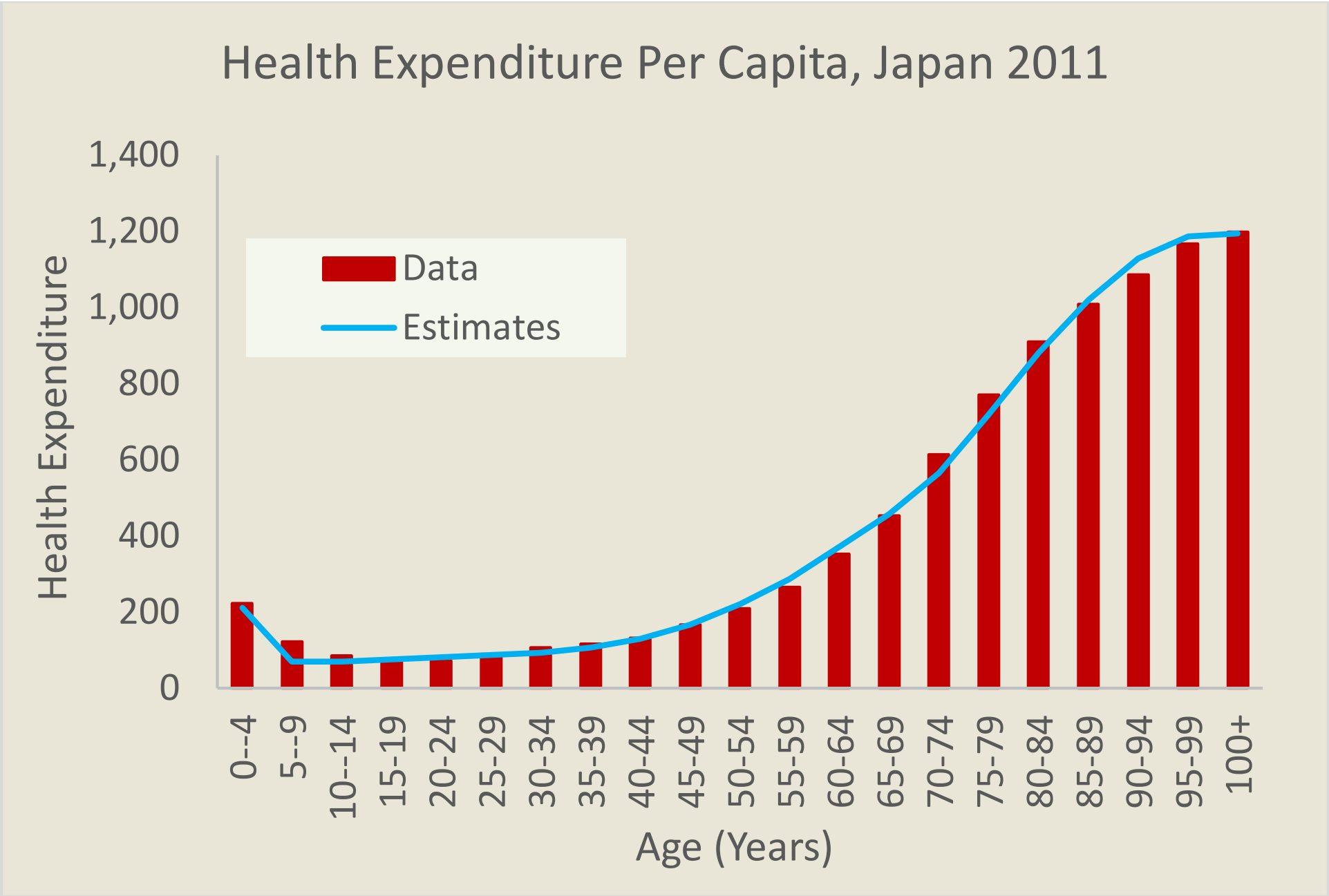

Fig. 6. Health Expenditure per Capita in Japan in 2011 (red bars) and Estimates (cyan curve)

The estimates for the health expenditure in Japan in 2011 in 1000s of Yen are illustrated in Figure 6. Clustered columns represent the data sets in Japan currency and the cyan curve our estimates. The statistics for these estimates are already referred earlier obtained from the extended life table presented in figure 4. The approach provided verifies to applicability and importance of our theory and the methodological and applied tools provided.

## 4. Conclusions and Further Study

We have developed an easy to apply methodology and designed expanded life tables to estimate the health expenditure in a country based on the Life Tables of this country. The application in Japan is quite promising that analogous studies will be possible for other countries thus supporting the decision makers and Health Agencies to develop effective strategies to better allocate the existing funds and further to make better predictions of the future health trends. A project is in progress on estimating the Health Expenditure in a Country by using the health state methodology. This application in Japan is in the framework of this project. Further applications in 20 countries are in the last estimation stages. Related programs and supporting tools are available upon request.

Japan-b-2011

**Estimates of HLYL and HLE (C H Skiadas and C Skiadas, 2016).**
**Add data in this Excel file and do not change the supporting files Gompertz and Weibull as they run automatically**

**Manually complete the blue Excel part only for mx or add all the Life Table from the Human Mortality Database, HMD (recommended)**

| | | | | | | | | | | | | Healthy Life Years Lost (HLYL) | | | | |
|---|---|---|---|---|---|---|---|---|---|---|---|---|---|---|---|---|
| | | | | | | | | | | | | HLYL Gompertz | HLYL Weibull | HLYL from mx | HLYL from qx | HLYL Average |
| | | | | | | | | | | | Years of age | 9.87 | 9.89 | 10.09 | 9.50 | 9.84 |
| Year | Age | mx | qx | ax | lx | dx | Lx | Tx | ex | | HLYL | HLE | | | | |
| 2011 | 0 | 0.00237 | 0.00237 | 0.21 | 100000 | 237 | 99813 | 8271181 | 82.71 | | 9.8 | 72.87 | 0.00237 | 0.00237 | 0.00237 | 0.00237 |
| 2011 | 1 | 0.00042 | 0.00042 | 0.5 | 99763 | 42 | 99742 | 8171368 | 81.91 | | 9.8 | 72.07 | 0.00279 | 0.00279 | 0.00042 | 0.00042 |
| 2011 | 2 | 0.00028 | 0.00028 | 0.5 | 99721 | 27 | 99707 | 8071626 | 80.94 | | 9.8 | 71.10 | 0.00307 | 0.00307 | 0.00056 | 0.00056 |
| 2011 | 3 | 0.00023 | 0.00023 | 0.5 | 99694 | 23 | 99682 | 7971919 | 79.96 | | 9.8 | 70.13 | 0.0033 | 0.0033 | 0.00069 | 0.00069 |
| 2011 | 4 | 0.00018 | 0.00018 | 0.5 | 99671 | 18 | 99662 | 7872236 | 78.98 | | 9.8 | 69.16 | 0.00348 | 0.00348 | 0.00072 | 0.00072 |
| 2011 | 5 | 0.00016 | 0.00016 | 0.5 | 99653 | 16 | 99645 | 7772575 | 78 | | 9.8 | 68.19 | 0.00364 | 0.00364 | 0.0008 | 0.0008 |
| 2011 | 6 | 0.00016 | 0.00016 | 0.5 | 99637 | 16 | 99629 | 7672930 | 77.01 | | 9.8 | 67.21 | 0.0038 | 0.0038 | 0.00096 | 0.00096 |
| 2011 | 7 | 0.00014 | 0.00014 | 0.5 | 99621 | 14 | 99614 | 7573301 | 76.02 | | 9.8 | 66.23 | 0.00394 | 0.00394 | 0.00098 | 0.00098 |
| 2011 | 8 | 0.00011 | 0.00011 | 0.5 | 99607 | 11 | 99602 | 7473687 | 75.03 | | 9.8 | 65.26 | 0.00405 | 0.00405 | 0.00088 | 0.00088 |
| 2011 | 9 | 0.00012 | 0.00012 | 0.5 | 99596 | 12 | 99590 | 7374085 | 74.04 | | 9.8 | 64.28 | 0.00417 | 0.00417 | 0.00108 | 0.00108 |
| 2011 | 10 | 0.00011 | 0.00011 | 0.5 | 99584 | 11 | 99579 | 7274494 | 73.05 | | 9.7 | 63.31 | 0.00428 | 0.00428 | 0.0011 | 0.0011 |
| 2011 | 11 | 0.00011 | 0.00011 | 0.5 | 99573 | 11 | 99567 | 7174916 | 72.06 | | 9.7 | 62.35 | 0.00439 | 0.00439 | 0.00121 | 0.00121 |
| 2011 | 12 | 0.0001 | 0.0001 | 0.5 | 99562 | 10 | 99556 | 7075348 | 71.07 | | 9.7 | 61.38 | 0.00449 | 0.00449 | 0.0012 | 0.0012 |
| 2011 | 13 | 0.00013 | 0.00013 | 0.5 | 99551 | 13 | 99545 | 6975792 | 70.07 | | 9.7 | 60.41 | 0.00462 | 0.00462 | 0.00169 | 0.00169 |
| 2011 | 14 | 0.00015 | 0.00015 | 0.5 | 99538 | 15 | 99530 | 6876247 | 69.08 | | 9.6 | 59.45 | 0.00477 | 0.00477 | 0.0021 | 0.0021 |
| 2011 | 15 | 0.00018 | 0.00018 | 0.5 | 99523 | 18 | 99514 | 6776717 | 68.09 | | 9.6 | 58.49 | 0.00495 | 0.00495 | 0.0027 | 0.0027 |
| 2011 | 16 | 0.00025 | 0.00025 | 0.5 | 99505 | 25 | 99493 | 6677203 | 67.1 | | 9.6 | 57.53 | 0.0052 | 0.0052 | 0.004 | 0.004 |
| 2011 | 17 | 0.00029 | 0.00029 | 0.5 | 99480 | 29 | 99466 | 6577710 | 66.12 | | 9.5 | 56.58 | 0.00549 | 0.00549 | 0.00493 | 0.00493 |
| 2011 | 18 | 0.00037 | 0.00037 | 0.5 | 99451 | 37 | 99432 | 6478245 | 65.14 | | 9.5 | 55.64 | 0.00586 | 0.00586 | 0.00666 | 0.00666 |
| 2011 | 19 | 0.00036 | 0.00036 | 0.5 | 99414 | 36 | 99396 | 6378812 | 64.16 | | 9.5 | 54.70 | 0.00622 | 0.00622 | 0.00684 | 0.00684 |
| 2011 | 20 | 0.0004 | 0.0004 | 0.5 | 99378 | 40 | 99358 | 6279416 | 63.19 | | 9.4 | 53.77 | 0.00662 | 0.00662 | 0.008 | 0.008 |
| 2011 | 21 | 0.00045 | 0.00045 | 0.5 | 99339 | 45 | 99316 | 6180058 | 62.21 | | 9.4 | 52.83 | 0.00707 | 0.00707 | 0.00945 | 0.00945 |
| 2011 | 22 | 0.00049 | 0.00049 | 0.5 | 99294 | 48 | 99270 | 6080741 | 61.24 | | 9.3 | 51.91 | 0.00756 | 0.00756 | 0.01078 | 0.01078 |
| 2011 | 23 | 0.00052 | 0.00052 | 0.5 | 99246 | 51 | 99220 | 5981471 | 60.27 | | 9.3 | 50.99 | 0.00808 | 0.00808 | 0.01196 | 0.01196 |
| 2011 | 24 | 0.00053 | 0.00053 | 0.5 | 99195 | 53 | 99168 | 5882251 | 59.3 | | 9.2 | 50.07 | 0.00861 | 0.00861 | 0.01272 | 0.01272 |
| 2011 | 25 | 0.00051 | 0.00051 | 0.5 | 99141 | 50 | 99116 | 5783083 | 58.33 | | 9.2 | 49.15 | 0.00912 | 0.00912 | 0.01275 | 0.01275 |
| 2011 | 26 | 0.00051 | 0.00051 | 0.5 | 99091 | 51 | 99066 | 5683967 | 57.36 | | 9.1 | 48.23 | 0.00963 | 0.00963 | 0.01326 | 0.01326 |
| 2011 | 27 | 0.0005 | 0.0005 | 0.5 | 99040 | 50 | 99016 | 5584901 | 56.39 | | 9.1 | 47.32 | 0.01013 | 0.01013 | 0.0135 | 0.0135 |
| 2011 | 28 | 0.00052 | 0.00052 | 0.5 | 98991 | 52 | 98965 | 5485886 | 55.42 | | 9.0 | 46.40 | 0.01065 | 0.01065 | 0.01456 | 0.01456 |
| 2011 | 29 | 0.00056 | 0.00056 | 0.5 | 98939 | 56 | 98911 | 5386921 | 54.45 | | 9.0 | 45.49 | 0.01121 | 0.01121 | 0.01624 | 0.01624 |
| 2011 | 30 | 0.00057 | 0.00057 | 0.5 | 98883 | 56 | 98855 | 5288010 | 53.48 | | 8.9 | 44.59 | 0.01178 | 0.01178 | 0.0171 | 0.0171 |
| 2011 | 31 | 0.00058 | 0.00058 | 0.5 | 98827 | 57 | 98798 | 5189155 | 52.51 | | 8.8 | 43.68 | 0.01236 | 0.01236 | 0.01798 | 0.01798 |
| 2011 | 32 | 0.00061 | 0.00061 | 0.5 | 98770 | 61 | 98740 | 5090356 | 51.54 | | 8.8 | 42.78 | 0.01297 | 0.01297 | 0.01952 | 0.01952 |
| 2011 | 33 | 0.00061 | 0.00061 | 0.5 | 98709 | 60 | 98679 | 4991617 | 50.57 | | 8.7 | 41.88 | 0.01358 | 0.01358 | 0.02013 | 0.02013 |
| 2011 | 34 | 0.0007 | 0.0007 | 0.5 | 98649 | 69 | 98615 | 4892938 | 49.6 | | 8.6 | 40.98 | 0.01428 | 0.01428 | 0.0238 | 0.0238 |
| 2011 | 35 | 0.00072 | 0.00072 | 0.5 | 98580 | 71 | 98545 | 4794323 | 48.63 | | 8.6 | 40.08 | 0.015 | 0.015 | 0.0252 | 0.0252 |
| 2011 | 36 | 0.00075 | 0.00075 | 0.5 | 98509 | 74 | 98472 | 4695778 | 47.67 | | 8.5 | 39.19 | 0.01575 | 0.01575 | 0.027 | 0.027 |
| 2011 | 37 | 0.00085 | 0.00085 | 0.5 | 98435 | 84 | 98393 | 4597306 | 46.7 | | 8.4 | 38.30 | 0.0166 | 0.0166 | 0.03145 | 0.03145 |
| 2011 | 38 | 0.00086 | 0.00086 | 0.5 | 98351 | 84 | 98309 | 4498913 | 45.74 | | 8.3 | 37.42 | 0.01746 | 0.01746 | 0.03268 | 0.03268 |
| 2011 | 39 | 0.00096 | 0.00096 | 0.5 | 98267 | 94 | 98220 | 4400604 | 44.78 | | 8.2 | 36.54 | 0.01842 | 0.01842 | 0.03744 | 0.03744 |
| 2011 | 40 | 0.00105 | 0.00105 | 0.5 | 98172 | 103 | 98121 | 4302384 | 43.82 | | 8.2 | 35.66 | 0.01947 | 0.01947 | 0.042 | 0.042 |
| 2011 | 41 | 0.00113 | 0.00113 | 0.5 | 98069 | 111 | 98014 | 4204263 | 42.87 | | 8.1 | 34.80 | 0.0206 | 0.0206 | 0.04633 | 0.04633 |
| 2011 | 42 | 0.00126 | 0.00126 | 0.5 | 97958 | 123 | 97896 | 4106250 | 41.92 | | 8.0 | 33.94 | 0.02186 | 0.02186 | 0.05292 | 0.05292 |
| 2011 | 43 | 0.00132 | 0.00132 | 0.5 | 97835 | 129 | 97770 | 4008353 | 40.97 | | 7.9 | 33.08 | 0.02318 | 0.02318 | 0.05676 | 0.05676 |
| 2011 | 44 | 0.00151 | 0.00151 | 0.5 | 97705 | 148 | 97632 | 3910583 | 40.02 | | 7.8 | 32.22 | 0.02469 | 0.02469 | 0.06644 | 0.06644 |
| 2011 | 45 | 0.00157 | 0.00157 | 0.5 | 97558 | 153 | 97481 | 3812952 | 39.08 | | 7.7 | 31.37 | 0.02626 | 0.02626 | 0.07065 | 0.07065 |
| 2011 | 46 | 0.00173 | 0.00173 | 0.5 | 97405 | 168 | 97321 | 3715470 | 38.14 | | 7.6 | 30.53 | 0.02799 | 0.02799 | 0.07958 | 0.07958 |
| 2011 | 47 | 0.00186 | 0.00186 | 0.5 | 97237 | 181 | 97146 | 3618150 | 37.21 | | 7.5 | 29.69 | 0.02985 | 0.02985 | 0.08742 | 0.08742 |
| 2011 | 48 | 0.00205 | 0.00205 | 0.5 | 97056 | 199 | 96956 | 3521003 | 36.28 | | 7.4 | 28.86 | 0.0319 | 0.0319 | 0.0984 | 0.0984 |
| 2011 | 49 | 0.00226 | 0.00226 | 0.5 | 96857 | 219 | 96748 | 3424047 | 35.35 | | 7.3 | 28.04 | 0.03416 | 0.03416 | 0.11074 | 0.11074 |
| 2011 | 50 | 0.00251 | 0.0025 | 0.5 | 96638 | 242 | 96517 | 3327299 | 34.43 | | 7.2 | 27.22 | 0.03667 | 0.03666 | 0.1255 | 0.125 |
| 2011 | 51 | 0.00267 | 0.00267 | 0.5 | 96396 | 257 | 96268 | 3230782 | 33.52 | | 7.1 | 26.42 | 0.03934 | 0.03933 | 0.13617 | 0.13617 |
| 2011 | 52 | 0.003 | 0.00299 | 0.5 | 96139 | 288 | 95995 | 3134515 | 32.6 | | 7.0 | 25.61 | 0.04234 | 0.04232 | 0.156 | 0.15548 |
| 2011 | 53 | 0.00323 | 0.00322 | 0.5 | 95851 | 309 | 95697 | 3038519 | 31.7 | | 6.9 | 24.82 | 0.04557 | 0.04554 | 0.17119 | 0.17066 |
| 2011 | 54 | 0.00345 | 0.00344 | 0.5 | 95543 | 329 | 95378 | 2942822 | 30.8 | | 6.8 | 24.03 | 0.04902 | 0.04898 | 0.1863 | 0.18576 |
| 2011 | 55 | 0.00364 | 0.00363 | 0.5 | 95214 | 346 | 95041 | 2847444 | 29.91 | | 6.7 | 23.25 | 0.05266 | 0.05261 | 0.2002 | 0.19965 |
| 2011 | 56 | 0.004 | 0.00399 | 0.5 | 94868 | 379 | 94679 | 2752403 | 29.01 | | 6.5 | 22.47 | 0.05666 | 0.0566 | 0.224 | 0.22344 |
| 2011 | 57 | 0.00446 | 0.00445 | 0.5 | 94489 | 420 | 94279 | 2657724 | 28.13 | | 6.4 | 21.71 | 0.06112 | 0.06105 | 0.25422 | 0.25365 |
| 2011 | 58 | 0.00481 | 0.0048 | 0.5 | 94069 | 452 | 93843 | 2563445 | 27.25 | | 6.3 | 20.95 | 0.06593 | 0.06585 | 0.27898 | 0.2784 |
| 2011 | 59 | 0.00535 | 0.00533 | 0.5 | 93617 | 499 | 93368 | 2469602 | 26.38 | | 6.2 | 20.20 | 0.07128 | 0.07118 | 0.31565 | 0.31447 |
| 2011 | 60 | 0.00587 | 0.00586 | 0.5 | 93118 | 545 | 92845 | 2376234 | 25.52 | | 6.1 | 19.47 | 0.07715 | 0.07704 | 0.3522 | 0.3516 |
| 2011 | 61 | 0.00624 | 0.00622 | 0.5 | 92573 | 576 | 92285 | 2283389 | 24.67 | | 5.9 | 18.75 | 0.08339 | 0.08326 | 0.38064 | 0.37942 |
| 2011 | 62 | 0.00683 | 0.00681 | 0.5 | 91997 | 626 | 91684 | 2191104 | 23.82 | | 5.8 | 18.03 | 0.09022 | 0.09007 | 0.42346 | 0.42222 |
| 2011 | 63 | 0.0075 | 0.00747 | 0.5 | 91371 | 683 | 91030 | 2099420 | 22.98 | | 5.7 | 17.32 | 0.09772 | 0.09754 | 0.4725 | 0.47061 |
| 2011 | 64 | 0.00832 | 0.00829 | 0.5 | 90688 | 751 | 90313 | 2008391 | 22.15 | | 5.5 | 16.62 | 0.10604 | 0.10583 | 0.53248 | 0.53056 |
| 2011 | 65 | 0.00797 | 0.00794 | 0.5 | 89937 | 714 | 89580 | 1918078 | 21.33 | | 5.4 | 15.94 | 0.11401 | 0.11377 | 0.51805 | 0.5161 |
| 2011 | 66 | 0.00946 | 0.00942 | 0.5 | 89223 | 840 | 88803 | 1828498 | 20.49 | | 5.3 | 15.23 | 0.12347 | 0.12319 | 0.62436 | 0.62172 |
| 2011 | 67 | 0.01052 | 0.01047 | 0.5 | 88382 | 925 | 87920 | 1739696 | 19.68 | | 5.1 | 14.56 | 0.13399 | 0.13366 | 0.70484 | 0.70149 |
| 2011 | 68 | 0.01094 | 0.01088 | 0.5 | 87457 | 952 | 86981 | 1651776 | 18.89 | | 5.0 | 13.92 | 0.14493 | 0.14454 | 0.74392 | 0.73984 |
| 2011 | 69 | 0.01223 | 0.01215 | 0.5 | 86505 | 1051 | 85980 | 1564795 | 18.09 | | 4.8 | 13.26 | 0.15716 | 0.15669 | 0.84387 | 0.83835 |
| 2011 | 70 | 0.01302 | 0.01294 | 0.5 | 85454 | 1106 | 84901 | 1478815 | 17.31 | | 4.7 | 12.63 | 0.17018 | 0.16963 | 0.9114 | 0.9058 |
| 2011 | 71 | 0.01436 | 0.01426 | 0.5 | 84349 | 1203 | 83747 | 1393914 | 16.53 | | 4.5 | 11.99 | 0.18454 | 0.18389 | 1.01956 | 1.01246 |
| 2011 | 72 | 0.01558 | 0.01546 | 0.5 | 83146 | 1285 | 82503 | 1310166 | 15.76 | | 4.4 | 11.38 | 0.20012 | 0.19935 | 1.12176 | 1.11312 |
| 2011 | 73 | 0.01788 | 0.01772 | 0.5 | 81860 | 1450 | 81135 | 1227663 | 15 | | 4.2 | 10.77 | 0.218 | 0.21707 | 1.30524 | 1.29356 |
| 2011 | 74 | 0.0192 | 0.01902 | 0.5 | 80410 | 1529 | 79645 | 1146528 | 14.26 | | 4.1 | 10.18 | 0.2372 | 0.23609 | 1.4208 | 1.40748 |
| 2011 | 75 | 0.02194 | 0.02171 | 0.5 | 78881 | 1712 | 78025 | 1066883 | 13.53 | | 3.9 | 9.61 | 0.25914 | 0.2578 | 1.6455 | 1.62825 |
| 2011 | 76 | 0.02431 | 0.02402 | 0.5 | 77169 | 1853 | 76242 | 988858 | 12.81 | | 3.8 | 9.05 | 0.28345 | 0.28182 | 1.84756 | 1.82552 |
| 2011 | 77 | 0.02744 | 0.02707 | 0.5 | 75315 | 2039 | 74296 | 912616 | 12.12 | | 3.6 | 8.52 | 0.31089 | 0.30889 | 2.11288 | 2.08439 |
| 2011 | 78 | 0.031 | 0.03053 | 0.5 | 73277 | 2237 | 72158 | 838321 | 11.44 | | 3.4 | 8.00 | 0.34189 | 0.33942 | 2.418 | 2.38134 |
| 2011 | 79 | 0.03478 | 0.03418 | 0.5 | 71040 | 2428 | 69825 | 766162 | 10.79 | | 3.3 | 7.52 | 0.37667 | 0.3736 | 2.74762 | 2.70022 |
| 2011 | 80 | 0.04006 | 0.03927 | 0.5 | 68611 | 2695 | 67264 | 696337 | 10.15 | | 3.1 | 7.05 | 0.41673 | 0.41287 | 3.2048 | 3.1416 |
| 2011 | 81 | 0.04394 | 0.04299 | 0.5 | 65917 | 2834 | 64500 | 629073 | 9.54 | | 2.9 | 6.60 | 0.46067 | 0.45586 | 3.55914 | 3.48219 |
| 2011 | 82 | 0.05 | 0.04878 | 0.5 | 63083 | 3077 | 61544 | 564573 | 8.95 | | 2.8 | 6.19 | 0.51067 | 0.50464 | 4.1 | 3.99996 |
| 2011 | 83 | 0.05537 | 0.05388 | 0.5 | 60005 | 3233 | 58389 | 503029 | 8.38 | | 2.6 | 5.79 | 0.56604 | 0.55852 | 4.59571 | 4.47204 |
| 2011 | 84 | 0.06424 | 0.06224 | 0.5 | 56772 | 3534 | 55005 | 444640 | 7.83 | | 2.4 | 5.42 | 0.63028 | 0.62076 | 5.39616 | 5.22816 |
| 2011 | 85 | 0.07262 | 0.07007 | 0.5 | 53239 | 3730 | 51373 | 389635 | 7.32 | | 2.2 | 5.08 | 0.7029 | 0.69083 | 6.1727 | 5.95595 |
| 2011 | 86 | 0.0807 | 0.07757 | 0.5 | 49508 | 3840 | 47588 | 338262 | 6.83 | | 2.1 | 4.68 | 0.7836 | 0.7684 | 6.9402 | 6.67102 |
| 2011 | 87 | 0.08911 | 0.08531 | 0.5 | 45668 | 3896 | 43720 | 290674 | 6.36 | | 2.0 | 4.34 | 0.87271 | 0.85371 | 7.75257 | 7.42197 |
| 2011 | 88 | 0.09949 | 0.09478 | 0.5 | 41772 | 3959 | 39792 | 246954 | 5.91 | | 1.9 | 4.01 | 0.9722 | 0.94849 | 8.75512 | 8.34064 |
| 2011 | 89 | 0.11349 | 0.10739 | 0.5 | 37813 | 4061 | 35782 | 207161 | 5.48 | | 1.8 | 3.71 | 1.08569 | 1.05588 | 10.1006 | 9.55771 |
| 2011 | 90 | 0.1266 | 0.11906 | 0.5 | 33752 | 4019 | 31743 | 171379 | 5.08 | | 1.7 | 3.43 | 1.21229 | 1.17494 | 11.394 | 10.7154 |
| 2011 | 91 | 0.14687 | 0.13683 | 0.5 | 29733 | 4068 | 27699 | 139636 | 4.7 | | 1.5 | 3.17 | 1.35916 | 1.31177 | 13.3652 | 12.4515 |
| 2011 | 92 | 0.15874 | 0.14706 | 0.5 | 25665 | 3774 | 23778 | 111937 | 4.36 | | 1.4 | 2.94 | 1.5179 | 1.45883 | 14.6041 | 13.5295 |
| 2011 | 93 | 0.18069 | 0.16572 | 0.5 | 21891 | 3628 | 20077 | 88159 | 4.03 | | 1.3 | 2.72 | 1.69859 | 1.62455 | 16.8042 | 15.412 |
| 2011 | 94 | 0.20033 | 0.18209 | 0.5 | 18263 | 3326 | 16600 | 68082 | 3.73 | | 1.2 | 2.53 | 1.89892 | 1.80664 | 18.831 | 17.1165 |
| 2011 | 95 | 0.22363 | 0.20114 | 0.5 | 14937 | 3004 | 13435 | 51482 | 3.45 | | 1.1 | 2.36 | 2.12255 | 2.00778 | 21.2449 | 19.1083 |
| 2011 | 96 | 0.24876 | 0.22124 | 0.5 | 11933 | 2640 | 10613 | 38047 | 3.19 | | 1.0 | 2.20 | 2.37131 | 2.22902 | 23.881 | 21.239 |
| 2011 | 97 | 0.27583 | 0.2424 | 0.5 | 9293 | 2253 | 8167 | 27434 | 2.95 | | 0.9 | 2.07 | 2.64714 | 2.47142 | 26.7555 | 23.5128 |
| 2011 | 98 | 0.3048 | 0.26449 | 0.5 | 7040 | 1862 | 6109 | 19267 | 2.74 | | 0.8 | 1.95 | 2.95194 | 2.73591 | 29.8704 | 25.92 |
| 2011 | 99 | 0.33554 | 0.28734 | 0.5 | 5178 | 1488 | 4434 | 13158 | 2.54 | | 0.7 | 1.85 | 3.28748 | 3.02325 | 33.2185 | 28.4467 |
| 2011 | 100 | 0.36791 | 0.31075 | 0.5 | 3690 | 1147 | 3117 | 8724 | 2.36 | | 0.6 | 1.75 | 3.65539 | 3.334 | 36.791 | 31.075 |
| 2011 | 101 | 0.40168 | 0.3345 | 0.5 | 2544 | 851 | 2118 | 5607 | 2.2 | | 0.5 | 1.68 | 4.05707 | 3.6685 | 40.5697 | 33.7845 |
| 2011 | 102 | 0.43658 | 0.35835 | 0.5 | 1693 | 607 | 1389 | 3488 | 2.06 | | 0.4 | 1.62 | 4.49365 | 4.02685 | 44.5312 | 36.5517 |
| 2011 | 103 | 0.47229 | 0.38206 | 0.5 | 1086 | 415 | 879 | 2099 | 1.93 | | 0.4 | 1.57 | 4.96594 | 4.40891 | 48.6459 | 39.3522 |
| 2011 | 104 | 0.50844 | 0.40538 | 0.5 | 671 | 272 | 535 | 1220 | 1.82 | | 0.3 | 1.53 | 5.47438 | 4.81429 | 52.8778 | 42.1595 |
| 2011 | 105 | 0.54465 | 0.42808 | 0.5 | 399 | 171 | 314 | 685 | 1.72 | | 0.2 | 1.49 | 6.01903 | 5.24237 | 57.1883 | 44.9484 |
| 2011 | 106 | 0.58054 | 0.44994 | 0.5 | 228 | 103 | 177 | 372 | 1.63 | | 0.2 | 1.46 | 6.59957 | 5.69231 | 61.5372 | 47.6936 |
| 2011 | 107 | 0.61572 | 0.47079 | 0.5 | 126 | 59 | 96 | 195 | 1.55 | | 0.1 | 1.44 | 7.21529 | 6.1631 | 65.882 | 50.3745 |
| 2011 | 108 | 0.64985 | 0.49048 | 0.5 | 66 | 33 | 50 | 99 | 1.49 | | 0.1 | 1.43 | 7.86514 | 6.65358 | 70.1838 | 52.9718 |
| 2011 | 109 | 0.6826 | 0.50891 | 0.5 | 34 | 17 | 25 | 49 | 1.43 | | 0.0 | 1.40 | 8.54774 | 7.16249 | 74.4034 | 55.4712 |
| 2011 | 110+ | 0.71372 | 1 | 1.4 | 17 | 17 | 23 | 23 | 1.4 | | 0.0 | 1.40 | | | | |

Fig. 7. Full Life Table and HLYL and HLE estimates